\documentclass[jap, aip, floatfix,reprint,twocolumn,superscriptaddress,showpacs,titlepage,amssymb,amsmath]{revtex4-1}

\usepackage{graphicx, float, times, epsfig, color, mathtools}
\def\be{\begin{equation}}
\def\ee{\end{equation}}
\def\bea{\begin{eqnarray}}
\def\eea{\end{eqnarray}}
\def\etal{{\it et al.}}
\def\deg{$^{\circ}$}
\def\asih{{\it a}-Si:H}
\def\asi{{\it a}-Si}

\begin{document} 
\title{Nearly defect-free dynamical models of disordered solids: The case of amorphous silicon}
\author{Raymond Atta-Fynn} 
\email{attafynn@uta.edu}
\affiliation{Department of Physics, University of Texas at Arlington, Arlington, TX 76019}

\author{Parthapratim Biswas}
\email{partha.biswas@usm.edu}
\affiliation{Department of Physics and Astronomy, 
The University of Southern Mississippi, Hattiesburg, MS 39406}

\begin{abstract}
It is widely accepted in the materials modeling community that defect-free
realistic networks of amorphous silicon cannot be prepared 
by quenching from a molten state of silicon using classical or {\it ab initio} 
molecular-dynamics (MD) simulations.  In this work, we address this 
long-standing problem by producing nearly defect-free ultra-large models
of amorphous silicon, consisting of up to half-a-million atoms, 
using classical molecular-dynamics simulations.  The structural, topological, 
electronic, and vibrational properties of the models are presented and 
compared with experimental data.  A comparison of the models 
with those obtained from using the modified Wooten-Winer-Weaire 
bond-switching algorithm shows that the models are on par 
with the latter, which were generated via event-based total-energy 
relaxations of atomistic networks in the configuration space.  
The MD models produced in this work represent the highest quality 
of amorphous-silicon networks so far reported in the literature 
using molecular-dynamics simulations. 
\end{abstract}

\maketitle 
\section{Introduction}

Silicon continues to play a major role in the technological 
revolution of the 21st century. The recent developments of 
high-efficiency silicon-based heterojunction solar cells~\cite{Taguchi_2014, Mishima_2011} and 
two-qubit quantum logic gates in silicon,~\cite{Veldhorst} with 
applications to quantum computation, are indicative of its lasting 
importance for future technology.  Unlike the crystalline state, the 
amorphous state of silicon is characterized by the presence of 
disorder in the radial and bond-angle distributions, along with a 
distinctly different topological (dis)ordering from its 
crystalline counterpart.  
Although a number of models based on the presence of small clusters 
of ordered materials -- variously known as the paracrystalline 
model,~\cite{Hoseman, para} the significant structure theory,~\cite{Walter} 
and the crystalline hypothesis~\cite{Bartenev}-- have been proposed 
from time to time, a CRN model appears to be conceptually simple 
and representative of most of the characteristic properties of 
{\asi} and {\asih}, observed in experiments. It is now widely 
accepted that the structure of amorphous silicon 
can be described fairly accurately by the continuous-random-network 
(CRN) model of Zachariasen.~\cite{crn}
The chemistry of silicon demands that an ideal CRN model 
of {\asi} should satisfy the following properties: (i) 
every Si atom in the network must be bonded to four 
neighboring Si atoms; (ii) the network must exhibit a 
well-defined short-range order and possibly an 
intermediate-range order, the former being characterized by 
narrow bond-length and bond-angle distributions (e.g., 
the average bond angle and its variance should be 
close to 109.47{\deg} and 9{\deg}--11{\deg}, 
respectively); (iii) any deviation from the ideal 4-fold coordination 
of the atoms must be as minimal as possible, preferably 
1--100 in 10$^5$ atoms, so that the resulting electronic, optical, and vibrational 
properties, obtained from such a model of {\asi}, are in agreement 
with experimental data.  Following Weaire and Thorpe,~\cite{WT_1971} 
and Heine,~\cite{Heine_1971} it can be shown, by employing 
a simple tight-binding Hamiltonian, that an atomistic model 
with these properties exhibits a 
gap in the electronic spectrum.  Despite these simple geometrical 
properties, atomistic modeling of defect-free {\asi} 
networks, using molecular-dynamics simulations, has been 
proved to be particularly challenging and, at present, no 
defect-free large molecular-dynamics models of {\asi} 
exist in the literature to our knowledge. 

The current approaches to structural modeling of {\it a}-Si can be broadly 
classified into three categories.  The first and foremost is based on Monte Carlo 
simulations, using the so-called bond-switching 
algorithm of Wooten, Winer and Weaire~\cite{W3, dtw} (W3). 
Here, one starts from a disordered silicon crystal with 
100\% 4-fold coordinated atoms and 
introduces a series of bond switches in the network, 
which are followed by total-energy relaxations using the 
Keating potential.~\cite{keating}
The bond switching between a pair of atoms changes the 
network topology, but keeps the atomic-coordination number 
constant, by incorporating mostly 5- and 7-member rings 
in the network.  A repeated application of bond switching, followed by 
total-energy relaxations, drives the system stochastically on 
the potential-energy surface (PES) from one local minimum 
to another over the energy barriers on the PES.  In order for the system to be able to 
explore the relevant part of the configuration space, which 
is associated with topologically-distinct 
configurations than a crystal, it is necessary to 
conduct a minimal number of bond switches so that 
the system can escape from the initial (crystalline) 
state.  An efficient implementation of the algorithm was presented by Barkema 
and Mousseau,\cite{BM} where the method was modified to start 
from a random configuration and introducing local relaxations, 
upon bond switching, followed by intermittent total-energy 
relaxations of the network. 
The method is capable of generating 
large 100\% defect-free {\asi} networks with a very narrow 
bond-angle distribution.~\cite{vink2} 

Molecular dynamics simulations, on the other hand,  provide 
an alternative route, where one attempts to generate {\asi} 
configurations by quenching from a molten state of Si at high temperature. 
Starting from a random configuration, with a mass density close to the 
experimental density~\cite{Custer1994} of 2.25 g/cm$^3$ for {\asi}, 
the temperature of the system is increased well above the melting point 
at 1687 K of $c$-Si. After thermalization, the temperature of 
the system is gradually reduced, typically at the rate of 5--10 K/ps, 
until the final temperature decreases to 300 K. The approach is 
commonly known as `quench-from-the-melt' and it is particularly 
effective for amorphous solids having 
strong glass-forming ability.  
It has been observed that {\asi} models obtained from the 
melt-quench approach contain a high density ($\ge$ 5\%) of 
coordination defects, namely 3-fold and 5-fold coordinated 
atoms, which is at variance with the dangling-bond density 
(i.e., 3-fold coordinated atoms) estimated from electron 
spin resonance (ESR) measurements.~\cite{ESR} Such a high 
density of coordination defects adversely affects the 
optoelectronic properties of the material, which renders MD 
models unsuitable for predictive studies of electronic and 
optical properties of {\asi} and {\asih}.  The apparent failure of 
melt-quench approaches to produce satisfactory models 
of {\asi} is not particularly surprising in view of the 
fact that {\asi} is not a glass and that it is generally 
prepared in laboratories by vapor deposition on a cold 
substrate or by similar methods.  The lack of {\it glassy} behavior in {\asi} can be qualitatively
understood from Phillips' constraint-counting approach~\cite{Phillips_1979}
and Thorpe's rigidity-percolation concept.~\cite{Thorpe_1983} In the latter,
the glassy nature of a disordered solid was shown to be
connected with the mean coordination number of the atoms in
a network. Specifically, a disordered network with a mean coordination
number $r_{\text{av}} < 2.4$ behaves as a {\it polymeric glass}, whereas networks
with $r_{\text{avg}} > 2.4$, such as {\asi} with $r_{\text{avg}} \approx 4$,
behave like rigid {\it amorphous solids}.
These observations, along with a high cooling rate and a rather 
short total simulation time, appear to indicate 
that realistic models of {\asi} is unlikely to be produced 
by molecular-dynamics simulations.  
Thus, atomistic modeling of {\asi} using direct MD simulations 
has attracted little attention in recent years~\cite{vasp_asi} 
and a great majority of MD studies on {\asi} were conducted 
in the past decades.\cite{justo,car,dave,cooper2000,kim,stich,klein,kluge,wd} 

In the last decade, a number of hybrid approaches were developed 
that ushered in a new direction for modeling complex materials 
using information paradigm.\cite{ecmr1, Pandey_SR2016, Prasai_2015, FPASS, AIRSS}
These methods employ prior knowledge of materials from experiments, often 
in conjunction with structural,~\cite{BiswasIOP2015}  
chemical~\cite{Pandey_PRB2015} and electronic~\cite{Prasai_2015} constraints, and 
combine the information with an appropriate  total-energy 
functional for structural determination of complex solids. 
Motivated by the Reverse Monte Carlo method~\cite{mac,gere,walt,biswas1,goodwin} 
and its drawback, where a three-dimensional structural model is generated 
by inverting one-dimensional experimental diffraction data, 
hybrid approaches go a step further by incorporating 
experimental data with a total-energy 
functional in search for structural solutions in an 
augmented solution space, so that the final structure 
is in agreement with both experiments and theory.  
Examples of such approaches include 
experimentally constrained molecular relaxation 
(ECMR),\cite{ecmr1, ecmr2} {\it ab initio} random structure 
searching (AIRSS),\cite{AIRSS} first-principles assisted structural 
solutions (FPASS),\cite{FPASS} and force-enhanced atomic relaxations 
(FEAR).\cite{Pandey_PRB2015, Pandey_SR2016} While the methods have 
achieved a remarkable success in determining structures of a variety 
of complex solids, none of the methods has been successful so far 
in producing high-quality, defect-free configurations of {\asi}. 
In this paper, we have addressed this problem and presented a solution by 
producing nearly defect-free large models of amorphous silicon using 
molecular-dynamics simulations, supported by experimental evidence 
from X-ray diffraction and Raman spectroscopy. 

The layout of the paper is as follows. In section II, we discuss 
a dynamical approach to modeling amorphous silicon using classical 
molecular-dynamics simulations in canonical and microcanonical 
ensembles, followed by {\it ab initio} total-energy relaxation 
within the framework of the density-functional theory. We successfully 
demonstrate that high-quality structural models of amorphous 
silicon, with defect concentrations as low as 0.7\% and 
root-mean-square deviations (RMS) of bond angles below 10{\deg}, can be 
obtained directly from ultra-long MD simulations that last 
for several tens of nanoseconds.  Section III discusses the results of our 
simulations with an emphasis on structural, electronic 
and vibrational properties of the models. We show that the 
results from the MD models are in excellent agreement with 
experimental data and that from the W3 models.  This is 
followed by the conclusions of our work in section IV.

\section{Computational Methodology}
\subsection{Dynamical approach to {\asi} using MD simulations}
Classical molecular-dynamics simulations were performed for 
several systems ranging from 300 to 400,000 atoms. 
The details of the simulations are as follows. 
(i) {\it Initial random configurations}:
Initial configurations of different sizes,  $N=$300, 1000, 5000, 25000, 400,000
Si atoms, were generated by randomly placing Si atoms in a cubic 
simulation box, subject to the constraint that the minimum separation between
each pair of atoms was 2.1 {\AA}. The length of the cubic 
simulation box was chosen such that the mass density 
matched with the experimental density~\cite{Custer1994} 
of {\asi} of 2.25 g/cm$^3$.  (ii) {\it MD simulations}:
MD simulations, using the initial random configurations as input,
were carried out in the canonical (NVT) and microcanonical (NVE) ensembles. 
The interatomic interaction between Si atoms were described 
using the modified Stillinger-Weber (SW) potential,~\cite{sw2,sw1} 
which is given by, 
\begin{equation}
 V(R^N)=\dfrac{1}{2}\sum_{i=1}^{N}\sum_{\substack{j=1 \\ (j\neq i)}}^{N}v_2(r_{ij}) 
      + \sum_{i=1}^{N}\sum_{\substack{j=1 \\ (j\neq i)}}^{N}\sum_{\substack{k=1\\(k\neq i)\\(k > j)}}^{N}v_3(\mathbf{r}_{ij},\mathbf{r}_{ik}),
\end{equation}
where $R^N$ indicates the atomic configuration and $v_2(r_{ij})$ is the two-body 
contribution to the potential energy given by,  
\begin{equation}
\begin{aligned}
v_2(r_{ij}) =  & \epsilon A\left[B\left(\dfrac{r_{ij}}{\sigma}\right)^{-p}-1\right]\exp\left(\dfrac{\sigma}{r_{ij}-a\sigma}\right)\\
               & \times \Theta\left(a\sigma - r_{ij}\right), 
\end{aligned}
\end{equation}
and $v_3(\mathbf{r}_{ij},\mathbf{r}_{ik})$ is the three-body contribution to the potential energy, 
\begin{equation}
\begin{aligned}
v_3(\mathbf{r}_{ij},\mathbf{r}_{ik})= & \epsilon\lambda\exp\left(\dfrac{\sigma\gamma}{r_{ij}-a\sigma}+ \dfrac{\sigma\gamma}{r_{ik}-a\sigma}\right)\\ 
                                      & \times \left(\cos(\hat{\mathbf{r}}_{ij}\cdot\hat{\mathbf{r}}_{ik})+\dfrac{1}{3}\right)^2\\
                                      & \times \Theta\left(a\sigma - r_{ij}\right)
\Theta\left(a\sigma - r_{ik}\right), 
\end{aligned}
\end{equation}
with $\Theta$ being the Heaviside step function. Here, we have used the 
modified potential parameters,  due to Vink {\it et al},~\cite{sw2} which are 
listed in Table~\ref{TAB1}. 

\begin{figure}[t!]
\includegraphics[clip,width=0.8\linewidth]{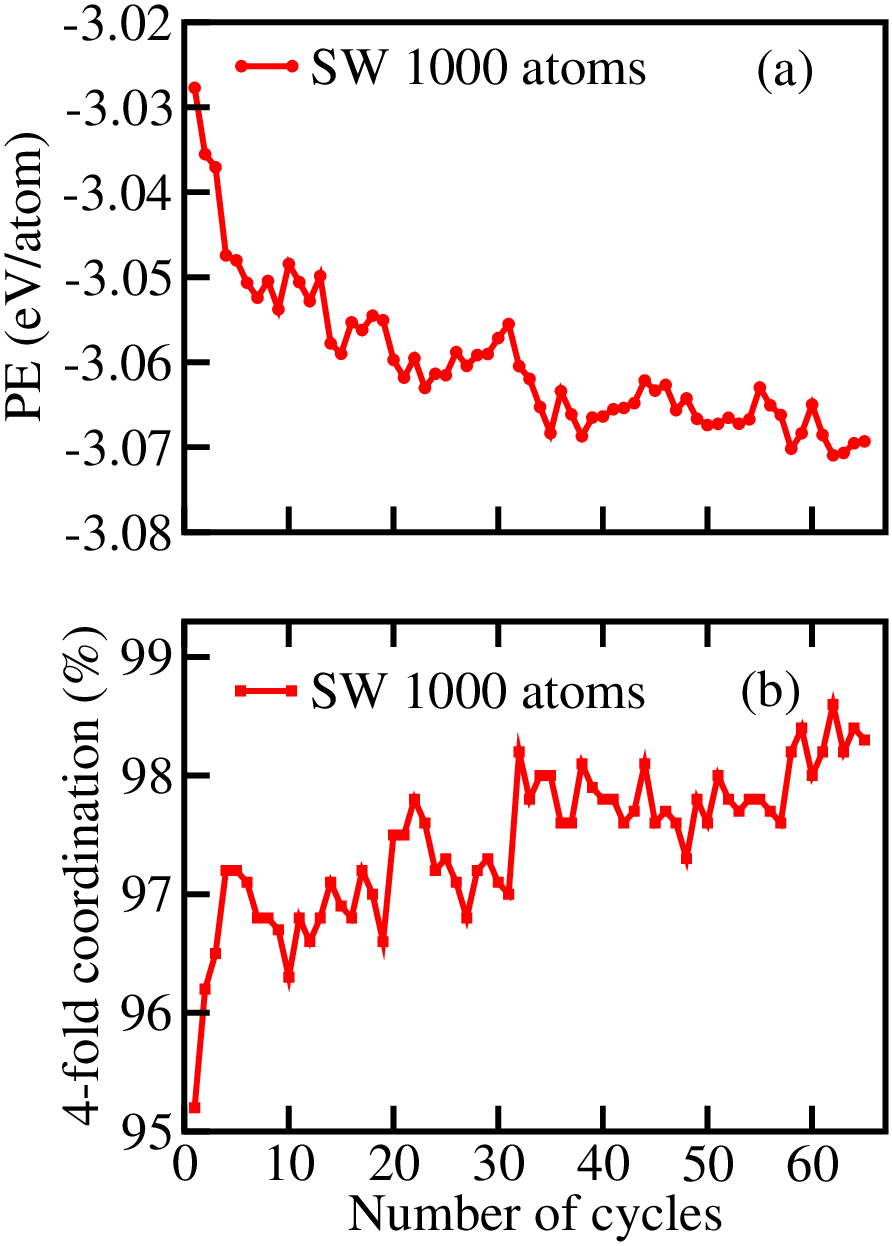}
\caption{\label{cool} 
(a) Time evolution of the potential energy per atom versus 
number of NVT-NVE cycles. In this illustrative plot, each cycle 
corresponds to a time period of 300 ps when the simulation 
temperature decreases from 1800 K to 300 K. (b) Evolution of the 
number of total atoms (in \%) having 4-fold coordination 
versus the number of NVT-NVE cycles. 
}
\end{figure}

The equations of motion were integrated, using the velocity-Verlet algorithm, 
with a time step of $\Delta t=1$ fs,  and a chain of 
Nos\'{e}-Hoover thermostats~\cite{chains,hoover,nose} was 
used to control the simulation temperature and the time evolution of the system 
in canonical ensembles.  Each system was initially 
equilibrated at 1800 K for 50 ps and then it was cooled to 300 K, over 
a total time period of 600 ps, by gradually decreasing the temperature 
in steps of 100 K with an average cooling rate of 2.5 K per ps.  At the 
end of the NVT-dynamics at 300 K, the system was 
subjected to evolve in a microcanonical ensemble (NVE) for an additional 
50 ps so that the total simulation time for a single NVT-NVE cycle 
from an initial temperature of 1800 K to the final temperature of 300 K was 700 ps.  
The purpose of the NVE run was to eliminate any possible artifacts that 
could originate from the thermostats and to collect structural 
configurations during the NVE dynamics, which were free from any 
thermal fluctuations.  Silicon 
configurations that satisfied a set of predefined criteria, such 
as those with a total potential energy and the density of coordination 
defects less than some prescribed 
values were collected during the 50-ps NVE run. On completion of the 
NVE run, the temperature of the system was increased to 1800 K 
to initiate the next NVT-NVE cycle.  In order for the system to 
be able to explore a considerable part of the phase space in 
searching for better amorphous configurations, the NVT-NVE cycle 
was repeated many times until there were no significant changes in 
the total energy and density of coordination defects.
In this study, we carried out 20-60 such cycles -- 
depending on the size of the systems -- for a duration of
700 ps per cycle, resulting in a total simulation time of 14-42 nanoseconds.
To demonstrate the efficacy of such NVT-NVE cycles, we have 
plotted in Fig.\,\ref{cool} the time evolution of the potential 
energy per atom and the fraction of the 4-fold coordinated 
atoms at the end of each cycle versus the number of cooling cycles.  
For the purpose of demonstration only, the data shown in 
Fig.\,\ref{cool} were generated using a high value of the 
cooling rate, 5 K/ps, for a 1000-atom SW model. 

\begin{table}
\caption{
\label{TAB1}
Modified Stillinger-Weber potential energy parameters due to Vink {\it et al}~\cite{sw2}.
}
\begin{ruledtabular}
\begin{tabular}{cccccccc}
      $\epsilon$ (eV) & $\lambda$ & $\sigma$ ({\AA}) & $\gamma$ & $A$ & $B$ & $a$ & $p$ \\
      1.64833  & 31.5 & 2.0951 & 1.20 & 7.049556277 & 0.6022245584 & 1.80 & 4 \\
  \end{tabular}
   \end{ruledtabular}
\end{table}

\noindent 
(iii) {\it Optimized configurations}: As mentioned before, a number of 
low-energy configurations with desired structural properties were collected 
during NVE evolution at the end of each cycle in step (ii).  The low-energy 
configurations were relaxed by minimizing the total energy using the modified SW potential 
with respect to the atomic positions. The energy minimization was carried out using the 
limited-memory BFGS algorithm.\cite{bfgs1,bfgs2}  
For comparison, W3 configurations\cite{W3} of sizes 300, 1000, and 
5000 atoms, each with a mass density of 2.25 g/cm$^3$, were also generated 
following the prescription of Barkema and Mousseau\cite{} (see upper 
panel of Table~\ref{TAB2}). 

\subsection{{\it Ab Initio} relaxation using density-functional theory}
The electronic structure of the SW-optimized models of {\asi} with 
300, 1000, and 5000 atoms was studied using the local-basis 
density-functional code 
{\sc Siesta}.~\cite{siesta}
Norm-conserving Troullier-Martins pseudopotentials,\cite{tm} which 
were factorized in the Kleinman-Bylander form, were employed in 
this work.\cite{kb} The 300- and 1000-atom models were optimized 
fully self-consistently using the Perdew-Burke-Ernzerhof (PBE)\cite{pbe} formulation of 
the generalized-gradient approximation (GGA) of the exchange-correlation energy, 
and double-$\zeta$ with polarization (DZP) basis functions. 
For the model with 5000 atoms, we resorted to the Harris-functional 
approach,~\cite{harris} which is based on the linearization of the 
Kohn-Sham equations in the density-functional theory.  We also employed 
single-$\zeta$ (SZ) basis functions and the local density approximation 
(LDA) to treat the exchange-correlation effects by using the 
Perdew-Zunger formulation of the LDA.~\cite{pz} 
Similar {\sc Siesta}-based {\it ab initio} total-energy 
optimizations of W3 models of same sizes and mass density 
were also carried out for comparison with the SW models (cf.\:lower panel 
of Table~\ref{TAB2}).
The vibrational density of states of 1000-atom SW and W3 models 
was computed, using the harmonic approximation, by constructing the 
dynamical matrix of the system. The latter was obtained by 
computing the average atomic force on each atom via numerical 
differentiation of the total energy, using an atomic 
displacement of $5\times 10^{-3}$ {\AA}.  The resulting 
dynamical matrix was diagonalized to obtain the vibrational 
frequencies and eigenvectors. 

\section{Results and Discussion}
\begin{figure}[t!]
\includegraphics[clip,width=0.8\linewidth]{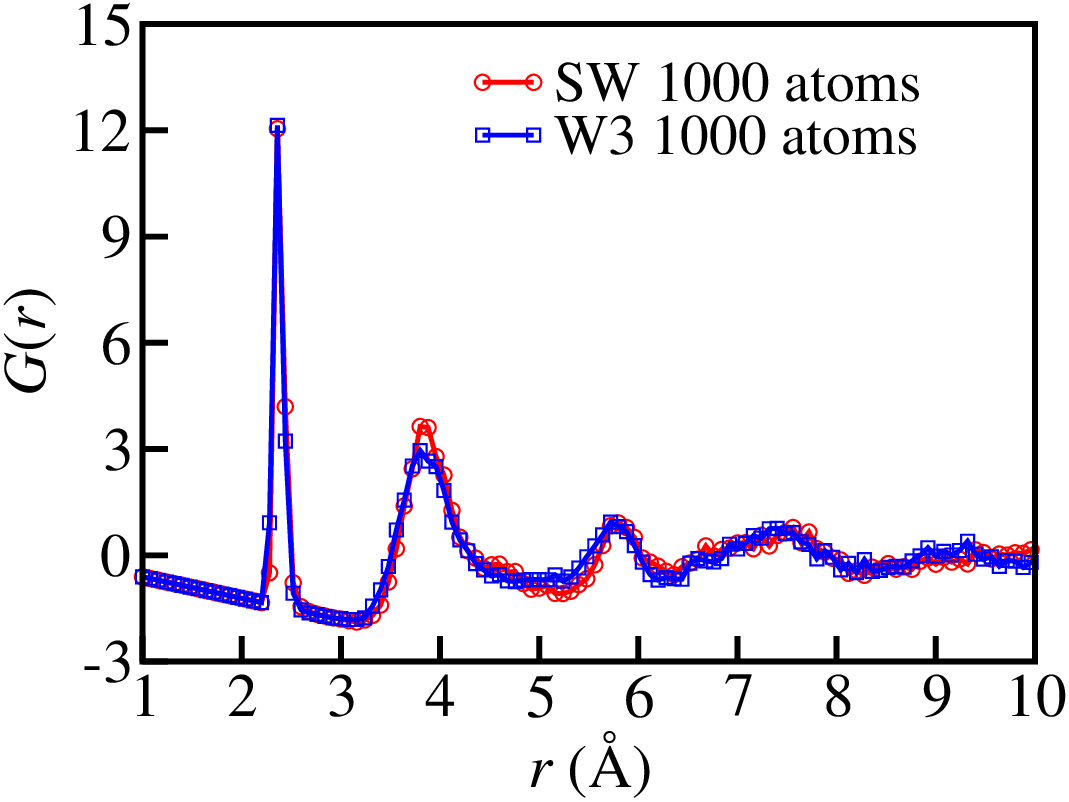}
\caption{\label{rdf}
(Color online)
The variation of the reduced pair-correlation 
functions ($G(r)$) with radial distances ($r$) 
for 1000-atom SW (red) and W3 (blue) models. 
}
\end{figure}

\begin{figure}[t!]
\includegraphics[clip,width=0.77\linewidth]{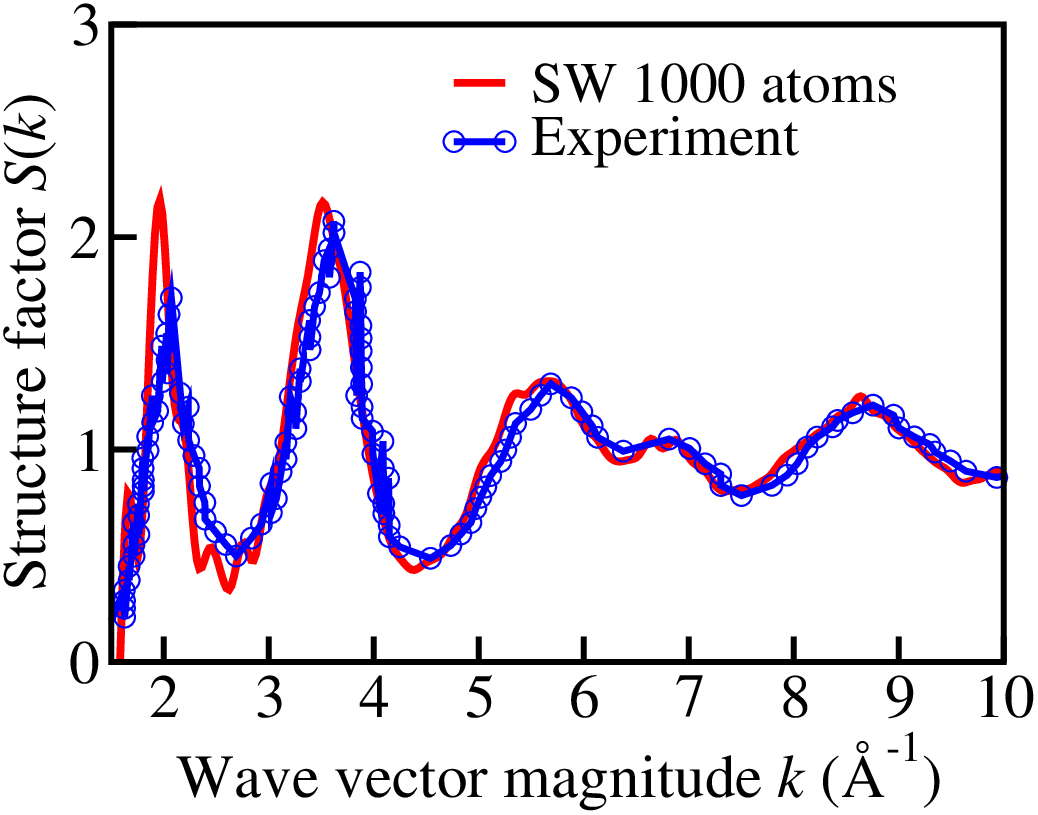}
\caption{\label{sk}
(Color online) 
Comparison of the static structure factors of {\asi}
from molecular-dynamics simulations (SW) and high-energy 
X-ray diffraction. The experimental data (blue) correspond to 
annealed samples of {\asi} from Laaziri {\etal}~\cite{laaziri}
}
\end{figure}

We begin by addressing the structural properties of the SW 
and W3 models.  As mentioned earlier, the quality of CRN 
models of {\asi} is primarily determined by: a) the radial 
distribution function; b) the bond-angle distribution (BAD) 
and its width; and c) the number of 4-fold coordinated atoms 
present in the network.  A high-quality CRN model must satisfy 
all these properties simultaneously. The bond-angle 
distribution must have an width, $\Delta \theta$, in the range 
9--11{\deg} and it must not have coordination defects more 
than 1 in 1000 atoms in order to be experimentally compliant. 
Currently, no MD models exist that can satisfy these 
requirements to our knowledge. In this section, we shall 
show that the MD models produced in our work, 
by combining NVT and NVE simulations discussed in 
Sec. IIA, closely satisfy all these requirements 
with a defect density of 0--1\% and a root-mean-square (RMS) 
bond-angle width of 9--10{\deg} for models with size up 
to 1000 atoms. 
Figure \ref{rdf} shows the reduced pair-correlation function (PCF), 
$G(r)$, versus radial distance ($r$) for {\sc Siesta}-optimized 
1000-atom SW and W3 models of {\asi}.  
The reduced PCF is defined as, $G(r) = 4\pi\rho\,r\,(g(r) - 1)$, where $\rho$ 
is the number density of Si atoms and $g(r)$ is the conventional 
pair-correlation function.   
The reduced PCFs from the W3 
and SW models matched closely in Fig.\,\ref{rdf}, showing an almost identical nature 
of two-body correlations between the atoms in the respective models. 
This observation is also reflected in Fig.\,\ref{sk}, where we 
have plotted the static structure factor,  
\be
S(k) =  \left\langle \frac{1}{N} \sum_i^N \sum_j^N 
\exp\left[\imath {\bf k} \cdot ({\bf r_i} - {\bf r_j})\right] \right\rangle, 
\label{sk-eq}
\ee
\noindent 
of a 1000-atom SW model, along with the experimental structure-factor data for annealed samples 
of {\asi}, due to Laaziri {\etal}\cite{laaziri} The symbol $\langle 
\rangle$ in Eq.\,\ref{sk-eq} indicates the rotational 
averaging of the wave-vector transfer ${\bf k}$ over a solid angle 
of $4\pi$, along 20000 different directions. 
A slightly high value of the first-sharp-diffraction peak (FSDP) for 
the SW model could be attributed partly to the use of the classical 
SW potential, producing a pronounced structural ordering, and in 
part to instrumental broadening associated with the collection of 
X-ray diffraction data.  A relatively slow cooling 
rate could also play a part in producing a somewhat more ordered 
structure.  We shall return to this point later to discuss the
topological order and ring statistics of the SW models.

Table \ref{TAB2} lists various structural properties of 
the SW and W3 models, from the average bond length and bond angle to the 
percentage of $n$-fold coordinated atom with $n$ = 3, 4, and 5. 
We have used a cutoff value of 2.8 {\AA} to calculate the 
coordination number of an atom, which corresponds to the 
first minimum of the PCF and it is practically identical 
for all models. The lower panel of Table \ref{TAB2} shows 
the results obtained from 
{\it ab initio} relaxations of the corresponding classical models. 
An examination of Table~\ref{TAB2} 
shows that the SW models contain a small fraction of coordination 
defects.  The percentage of 4-fold coordinated atoms can be seen 
to decrease only by 2\% as the system size increase from 300 
atoms to 400,000 atoms.  The defect 
density in the 300- and 1000-atom models was found to be no 
more than 1\% and, in each case, the RMS width ($\Delta \theta$) 
of the bond-angle distribution was observed to be less than 10{\deg}. To our 
knowledge, these are the best configurations so far obtained 
from any molecular-dynamics simulations to date. Although the 
models with 25000 and 400,000 Si atoms have a somewhat high 
defect concentration,  in the range (2.2--2.4)\%, the results 
can be readily appreciated in view of the large phase-space 
volume involved in the simulation of these ultra-large models.  
We should emphasize the fact that the MD models reported here were 
obtained from unconstrained atomic dynamics without using 
any coordination constraints unlike the W3 models, where a 
4-fold nearest-neighbor list was always maintained during 
simulation.

\begin{table}
\caption{\label{TAB2}
Structural properties of SW and W3 models.  $N$, $\langle \theta \rangle$,  
$\Delta\theta$,  $\langle n \rangle$, and $C_k$ indicate the 
number of atoms, the average bond length ({\AA}) and 
bond angle (degree), the RMS width of the bond angles (degree), the 
mean coordination number, and the percentage of $k$-fold 
coordinated atoms, respectively.  SW and W3 denote the results 
from molecular-dynamics simulations and the bond-switching algorithm 
of Wooten, Winer and Weaire, respectively.~\cite{W3, dtw}
}
   \begin{ruledtabular}
    \begin{tabular}{llccccccc}
       Model                          &
       $N$                            &
       $d$                            &
       $\langle \theta \rangle$       &
       $\Delta\theta$                 &
       $\langle n \rangle$            &
       $C_3$ (\%)                     &
       $C_4$ (\%)                     &
       $C_5$ (\%)                     \\
      \hline
      &\multicolumn{8}{c}{W3 and SW-MD Models}\\
      \hline
       SW  & 300     & 2.38  & 109.22 &  9.06 & 4 & 0.3 & 99.3   & 0.3 \\
       W3 & 300     & 2.36  & 109.26 &  9.74 & 4 & 0    & 100   & 0   \\
       SW  & 1000    & 2.39  & 109.19 &  9.69 & 4 & 0.7  & 99    & 0.3 \\
       W3 & 1000    & 2.37  & 109.25 &  9.06 & 4 & 0    & 100   & 0.0 \\
       SW  & 5000    & 2.39  & 109.18 &  9.62 & 4 & 1.0  & 98    & 1.0 \\
       W3 & 5000    & 2.36  & 109.22 &  9.84 & 4 & 0    & 100   & 0   \\
       SW & 25000  & 2.38  & 109.24  & 9.12 & 4  & 1.2 &  97.8  & 1.0 \\
       SW & 400000  & 2.39 & 109.22  & 9.46 & 4  & 1.2 &  97.6 & 1.2 \\
      \hline
      &\multicolumn{8}{c}{{\it Ab Initio}-relaxed W3 and SW-MD Models}\\
      \hline
       SW  & 300     & 2.38  & 109.20 &  9.44 & 4 & 0.3 & 99.3 & 0.3   \\
       W3 & 300     & 2.37  & 109.12 & 10.73 & 4 & 0    & 100   & 0   \\
       SW  & 1000    & 2.39  & 109.23 &  9.40 & 4 & 0.4  & 99    & 0.6 \\
       W3 & 1000    & 2.37  & 109.16 & 10.08 & 4 & 0    & 100   & 0.  \\
       SW  & 5000    & 2.39 & 109.19 & 9.49  & 4 & 1.0  & 98     & 1.0 \\
       W3 & 5000    & 2.37  & 109.11 &  10.71 & 4 & 0    & 100   & 0   \\
    \end{tabular}
   \end{ruledtabular}
\end{table}

\begin{figure}[tbp!]
\includegraphics[clip,width=0.75\linewidth]{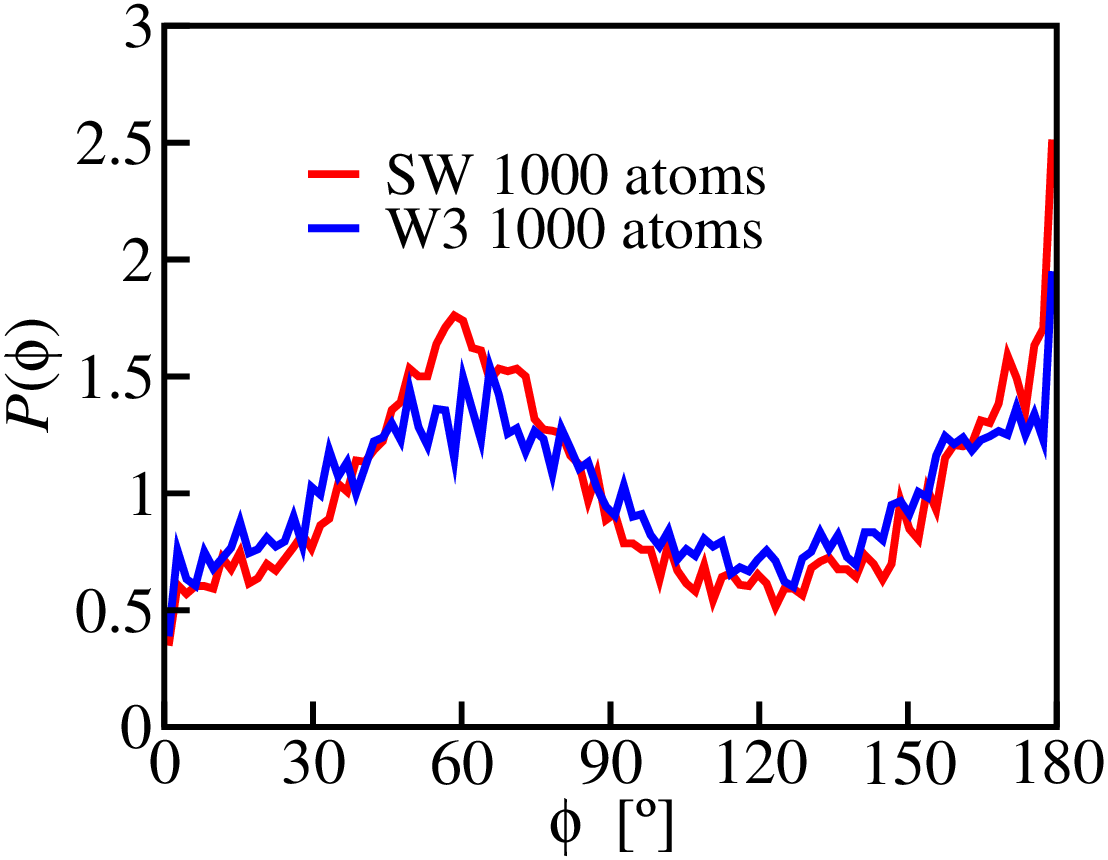}
\caption{\label{dihed}
(Color online)
The distribution ($P(\phi)$) of dihedral angles ($\phi$) in 
{\asi} for SW (red) and W3 (blue) models of size 1000 atoms. 
}
\end{figure}

Earlier we have mentioned that the PCF and the concentration of 
4-fold coordinated atoms alone are not sufficient to 
determine the quality of CRN models of {\asi}. For a 
high-quality CRN model, the distribution of bond angles must be 
sufficiently narrow, with an RMS width of the bond-angle distribution 
in the range 9--11{\deg}, as observed 
in X-ray~\cite{laaziri} and Raman transverse-optic (TO) 
peak measurements.~\cite{Beeman1985} 
Thus, in addition to the correct PCF, a CRN model of {\asi} must have 
appropriate values of $C_4$, $\langle \theta \rangle$, 
and $\Delta \theta$ so that the computed properties from the model 
are consistent with electron spin resonance,~\cite{ESR} X-ray diffraction,~\cite{laaziri} 
and Raman spectroscopy,~\cite{Beeman1985} respectively. 
It is notable that all the reported values of $\Delta \theta$ 
for the SW models, including the one with 400,000 Si atoms, 
in Table~\ref{TAB2} lie in the range of 9\deg--10\deg. 
The dihedral-angle distributions for the 1000-atom SW and W3 
models are shown in Fig.\,\ref{dihed}. The distributions are 
quite similar with a characteristic peak at 60{\deg}.  

Further characterization of SW and W3 models of {\asi} is 
possible by examining the site-averaged orientational 
order parameter (OOP), $Q_l$, due to Steinhardt {\etal}~\cite{Steinhardt}
In spherical polar coordinates, the site-projected OOP is defined as,  
\be
Q_l^i = \sqrt{\frac{4\pi}{2l+1} \sum_{m=-l}^{l} \left|\frac{1}{n_i}\sum_{j \in \{n_i\}}
Y_l^m(\theta({\bf r_{ij}}),\phi({\bf r_{ij}}))\right|^2}, 
\label{QL}
\ee
and the corresponding site-averaged value follows from, 
\[
Q_l  = \frac{1}{N}\sum_{i=1}^N Q_l^i.  \notag 
\] 

\begin{figure}[tbp!]
\includegraphics[clip,width=0.75\linewidth]{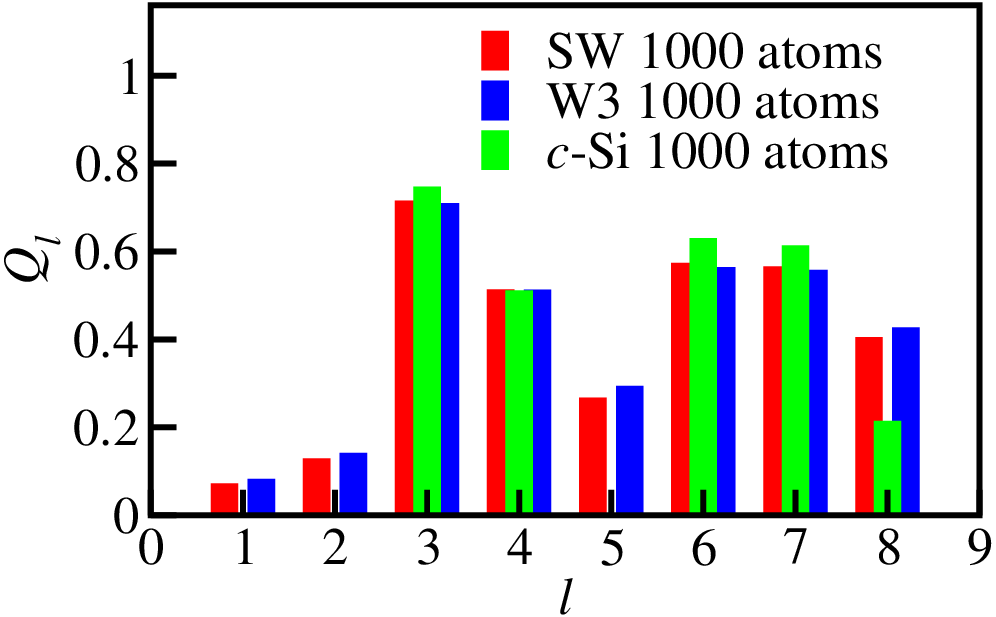}
\caption{\label{ql}
(Color online)
The site-averaged bond-orientational order parameter, $Q_l$, 
versus $l$ for SW (red), W3 (blue), crystalline silicon 
(green) networks. 
}
\end{figure}

\noindent 
In Eq.\,(\ref{QL}), $Q_l^i$ is the site-projected OOP associated 
with site $i$, whose value depends on the orientation of $n_i$ 
bonds that extend from site $i$ to the neighboring sites 
$j$. Here, $N$ is the system size, and $\theta$ and $\phi$ are 
the polar and azimuthal angles of a bond ${\mathbf r_{ij}}$, 
respectively. Figure \ref{ql} show the magnitude of $Q_l$, 
$l$ = 1 to $l$ = 8, for SW, W3, and crystalline networks, each 
consisting of 1000 atoms. A crystalline silicon network (in diamond structure) 
is characterized by a null or zero value of $Q_1$, $Q_2$, and $Q_5$. By 
contrast, SW and W3 models show a strong presence of $Q_5$ 
indicating the amorphous character of the networks. It is notable 
that $Q_l$ values for SW and W3 models practically match with each 
other, reflecting an identical nature of local environment as far 
as the orientational ordering of the atoms in the first coordination 
shell (of an atom) is concerned.

\begin{figure}[htbp!]
\includegraphics[clip,width=0.85\linewidth]{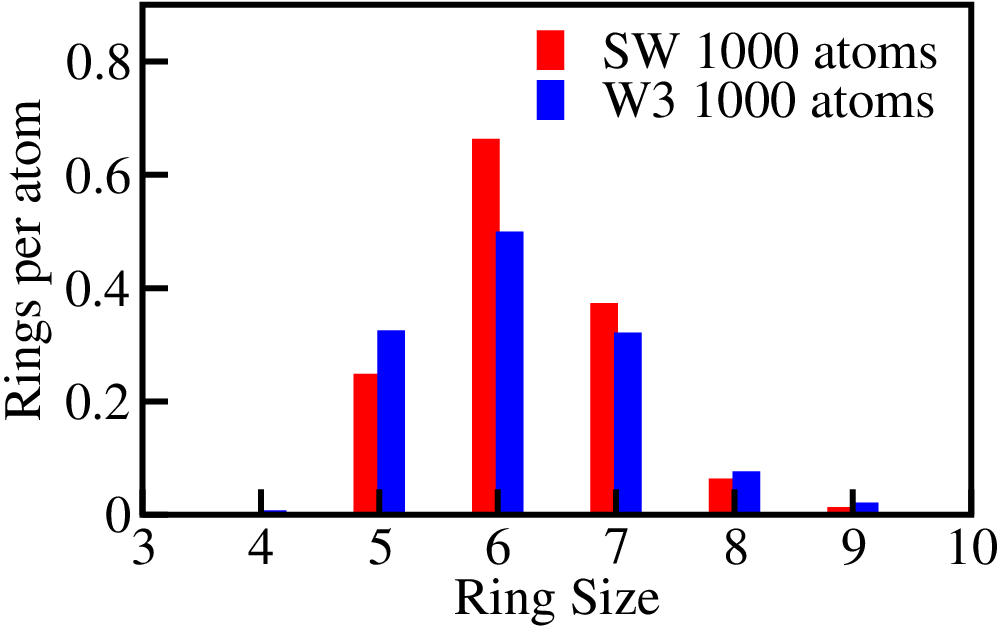}
\caption{\label{ring}
(Color online)
The distribution of irreducible rings/atom in {\sc Siesta}-relaxed 
1000-atom SW (red) and W3 (blue) models. A high value of 
6-member rings/atom in the SW model is indicative of a more 
topologically-ordered structure. See text for discussions. 
}
\end{figure}

Having addressed key structural properties of the SW models, we 
now examine their topological character in relation 
to the W3 models. Since the latter were generated by inducing, mostly, 
5- and 7-member rings in a network, it is instructive to examine 
to what extent the ring statistics from W3 models, obtained 
from an event-based bond-switching algorithm, compare with 
the same from SW models.
Toward that end, we have computed the 
number of $n$-member irreducible rings, from $n$=3 to $n$=10, and listed the 
values in Table \ref{TAB3}. 
Mathematically, an irreducible ring of size $n$ is defined as 
the shortest, self-avoiding, irreversible path, which 
starts and ends at the same atomic site in $n$ steps.  Here, 
irreducibility refers to the fact that such a ring cannot be 
partitioned further into a smaller set of rings without 
changing the topology of the path or circuit.  
Figure \ref{ring} presents a comparison of the ring-size distribution 
between an SW model and a W3 model of size 1000 atoms. In computing 
the irreducible ring-size distributions, we have not imposed 
periodic boundary conditions.  Since crystalline silicon, 
without any defects, is characterized by 
6-member rings only, the fraction of 6-member rings present 
in a network can be used as a measure of the degree of 
crystallinity of the network from a topological point of 
view.  Thus, as far as the ring statistics of the W3 and 
SW models are concerned in Fig.\,\ref{ring}, the latter can be 
described as having a somewhat pronounced topological 
ordering in comparison to the former. This observation 
is reflected in the FSDP of the SW models in 
Fig.\,\ref{sk}. 

\begin{table}
\caption{\label{TAB3}
Irreducible ring statistics (rings/per atom) for SW and W3 models.}
\begin{ruledtabular}
    \begin{tabular}{llccccccc}
       Model  & $N$ & 4 & 5 & 6 & 7 & 8 & 9 & 10 \\  
    \hline
      &\multicolumn{8}{c}{W3 and SW-MD Models}\\
      \hline
       SW  & 1000    & 0.001 & 0.245 & 0.662 & 0.377 & 0.063 & 0.011 & 0.003 \\
       W3 & 1000    & 0.006 & 0.324 & 0.496 & 0.315 & 0.074 & 0.021 & 0.001 \\
       SW  & 5000    & 0.002 & 0.297 & 0.782 & 0.453 & 0.077 & 0.017 & 0.001 \\
       W3 & 5000    & 0.020 & 0.391 & 0.565 & 0.374 & 0.121 & 0.022 & 0.003 \\
      \hline
      &\multicolumn{8}{c}{{\it Ab Initio}-relaxed W3 and SW-MD Models}\\
      \hline
       SW  & 1000    & 0.002 & 0.247 & 0.662 & 0.372 & 0.062 & 0.012 & 0.003 \\
       W3 & 1000    & 0.006 & 0.324 & 0.498 & 0.320 & 0.075 & 0.020 & 0.005 \\
       SW  & 5000    & 0.003 & 0.297 & 0.782 & 0.453 & 0.078 & 0.017 & 0.001 \\ 
       W3 & 5000    & 0.020 & 0.391 & 0.566 & 0.373 & 0.121 & 0.022 & 0.003 \\
    \end{tabular}
   \end{ruledtabular}
\end{table}

\begin{figure}[htbp]
\includegraphics[clip, width=0.8\linewidth]{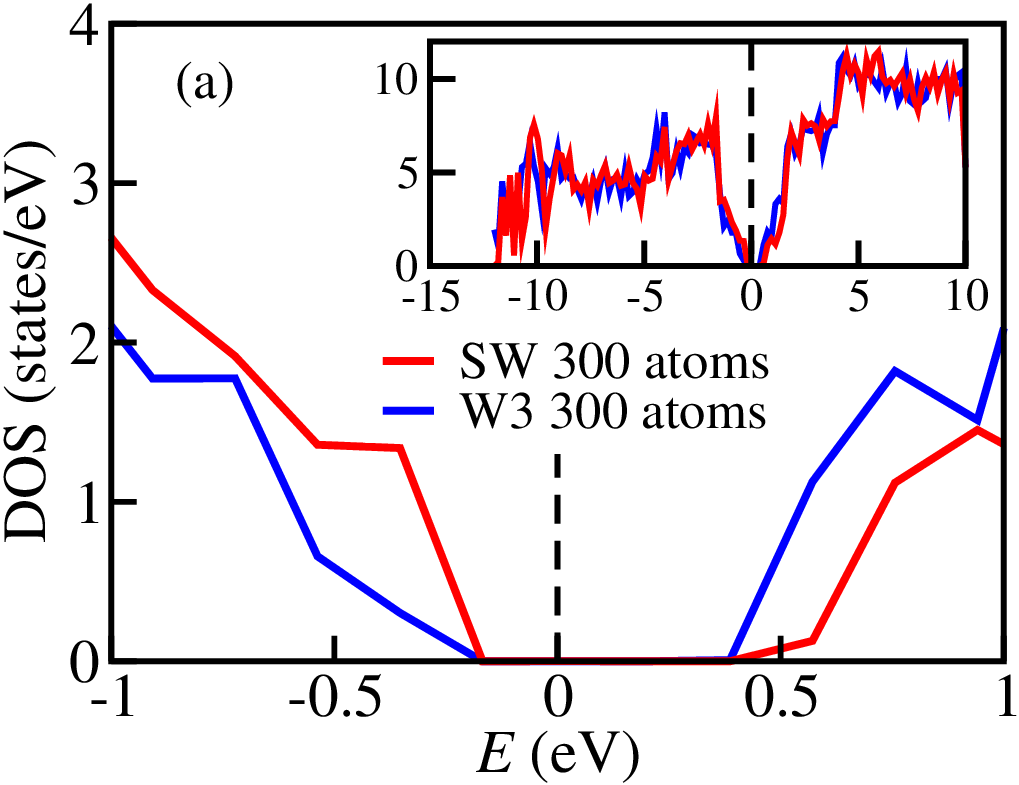}
\hspace*{-0.5cm}
\includegraphics[clip, width=0.85\linewidth]{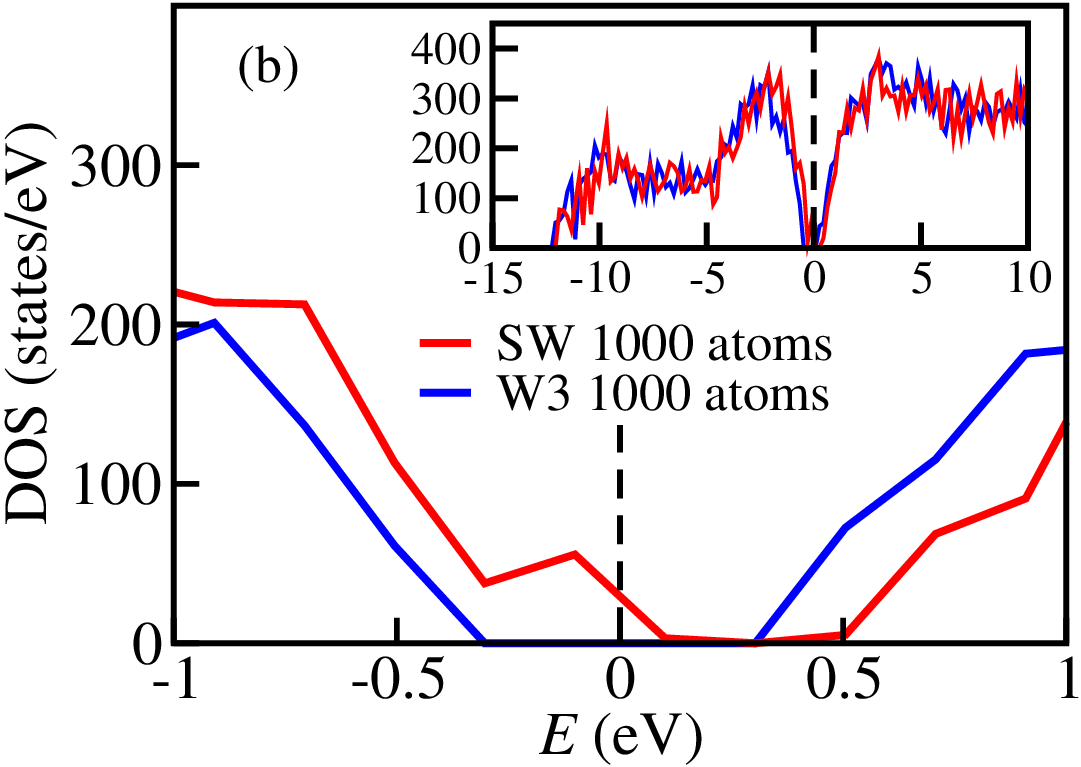}
\caption{\label{fig3}
(Color online) 
Depiction of the electronic densities of states (EDOS) of {\asi} from SW (red) 
and W3 (blue) models in the regions close to their band-gaps.  
The full EDOS for the 300- and 1000-atom models are 
shown as inset in (a) and (b), respectively.  
The Fermi level is indicated by a dashed vertical 
line at 0 eV. 
} 
\end{figure}

A good atomistic configuration of {\asi} must exhibit a 
gap in the electronic density of states (EDOS), irrespective 
of the method of preparation or modeling. The size of the 
gap depends on a number of factors, such as the density 
of coordination defects, the type of the defects (e.g., 
dangling and floating bonds), and the degree of disorder 
in the network. In general, a high density of 3-fold coordinated 
atoms or dangling bonds results in a large number of 
electronic states to appear near the Fermi level, which lead to 
a noisy or gap-less density of states. Figure~\ref{fig3} shows 
the EDOS for the 300- and 1000-atom SW and W3 models of {\asi}, 
with the Fermi level at 0 eV.  For the 300-atom SW model, we get 
a reasonably clean gap with two defects near the Fermi 
level, whereas the 1000-atom SW model produces a somewhat 
small gap due to 1\% coordination defects. At any rate, a 
band gap of size 0.6-0.8 eV has been realized 
in the SW models. A few defects states that are present 
in the vicinity of the band-gap region can be readily 
passivated by hydrogenation to further improve the 
quality of the MD models.  

\begin{figure}[t!]
\includegraphics[clip,width=0.8\linewidth]{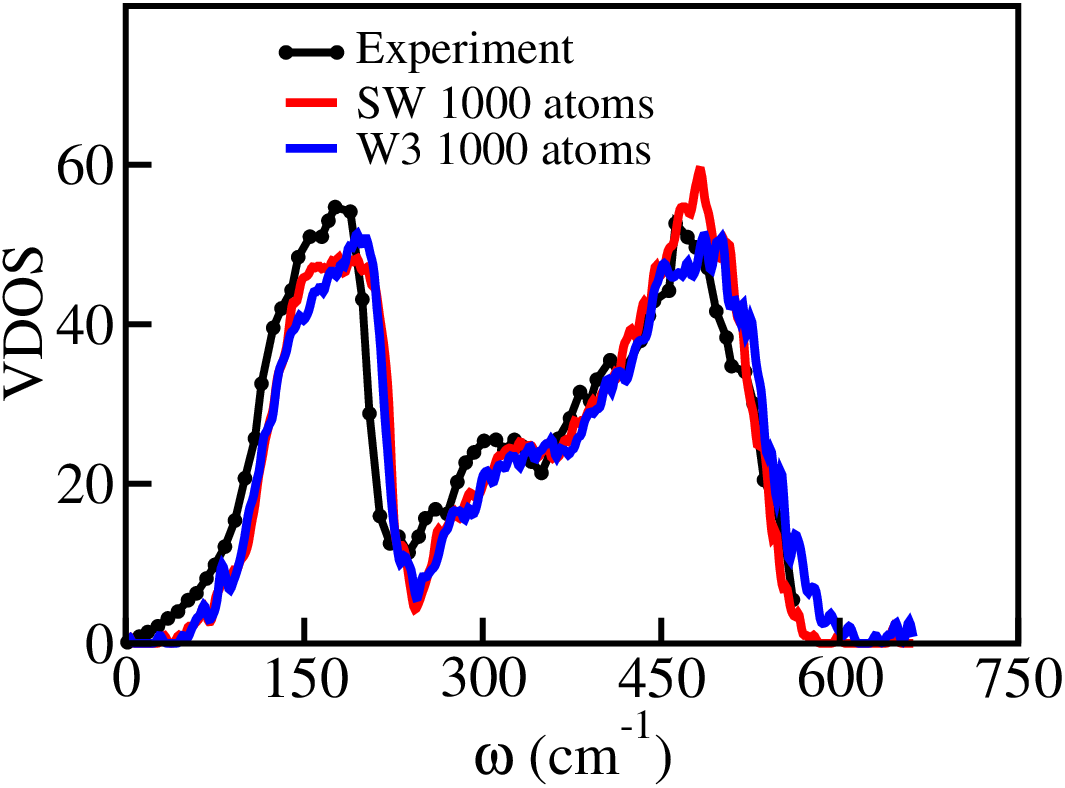}
\caption{\label{fig4}
(Color online) {\it Ab initio} vibrational densities of states 
from 1000-atom SW and W3 models.  Inelastic neutron-scattering 
experimental data, due to Kamitakahara {\etal},~\cite{Kamitakahara} 
are also included in the plot for comparison.
}
\end{figure}

Finally, a true atomistic model must reproduce correctly the 
vibrational density of states (VDOS). Since vibrational excitations 
are more sensitive to local atomic environment than their electronic 
counterparts, any minute changes in the local atomic structure of 
{\asi} can readily reflect on the VDOS at room 
temperature.~\cite{vib} The VDOS was computed by constructing the force-constant 
(FC) matrix, using the harmonic approximation, and diagonalizing 
the corresponding dynamical matrix. 
Figure~\ref{fig4} shows the VDOS of {\asi}, obtained from 
1000-atom SW and W3 models.  
To compare the results with experimental data, we have 
also plotted the experimental inelastic neutron-scattering 
data from Kamitakahara {\it et al.}~\cite{Kamitakahara}
The vibrational density of states from the 
SW and W3 models matches closely with the experimental 
data. 

\section{conclusions}

In this paper, we have addressed a long-standing problem of 
structural modeling of {\asi} using molecular-dynamics (MD) 
simulations. It is generally believed that, since {\asi} is not a 
glass, direct MD simulations of {\asi}, by quenching from 
a molten state of Si at high temperature, are not possible.  
A review of numerous MD studies on {\asi} appears to 
support this conjecture by noting that MD simulations tend 
to produce a high density of coordination defects, which is 
inconsistent with the dangling-bond density observed in {\asi} 
thin-films, by electron spin resonance 
measurements.  This observation is consistent with the fact 
that {\asi} samples are prepared in laboratories not by 
melt-quenching but by vapor deposition on a cold substrate 
or equivalent methods.  Despite these observations, in this 
paper we have shown conclusively that it is possible to 
simulate high-quality CRN models of amorphous silicon
by employing ultra-long molecular-dynamics simulations 
spanning several tens of nanoseconds.  The size of 
the models produced in this study ranges from 300 
atoms to 400,000 atoms and the corresponding 
defect density lies between 0.7\% to 2.4\%, respectively. 
All the MD models exhibit a very narrow bond-angle distribution, 
characterized by an average bond angle of $\sim$ 109.2 
and a root-mean-square (RMS) deviation 
of $\sim$ 9.4{\deg}--9.7{\deg}.  The latter is fully consistent with the value extracted from 
Raman TO peak measurements. Our results can be fully 
appreciated by noting that the 300-atom model has only a pair of 
defects -- one dangling bond and one floating bond -- and 
a bond-angle width of 9.44{\deg}, which exhibits a clean 
gap in the electronic density of states.
On the other hand, the largest model with 400,000 atoms 
has a defect density of 2.4\% only and an RMS width
of 9.46{\deg} in the bond-angle distribution. 

We conclude this section with the following observations. 
First, the availability of nearly defect-free MD 
models of {\asi} brings considerable advantages in 
studying {\asi} and {\asi}-related materials. A 
molecular-dynamical approach provides a natural route 
to produce atomistic models of {\asi}. 
The approach is simple and intuitive and it 
can be implemented easily and efficiently for large
models, both 
in serial and parallel computing environments.
Second, unlike the W3 method, a molecular-dynamical 
approach is free from any constraints. In the former, 
the system evolves from one configuration to another, 
by employing a limited set of bond switches, 
such that the coordination number of the atoms remains 
constant. 
While such bond reassignments can efficiently 
induce the requisite 5- and 7-member rings 
to produce the characteristic network topology 
of {\asi}, the system can only explore a coarse-grained 
configuration space due to the 
coordination constraint and a limited combination 
of bond switches. Thus, the topological character 
of the resulting amorphous networks from such an approach 
may not be necessarily identical to the 
ones obtained from unconstrained MD simulations.
This observation is reflected in the ring 
statistics obtained from the SW and W3 models. 
Since rings can play an important role in determining 
the stability and topological ordering of amorphous/disordered 
networks, it remains to be seen to what extent 
MD models and W3 models differ from each other and 
affect the material properties, involving higher-order 
correlation functions. Finally, the availability of large 
MD models will be particularly useful in studying 
silicon-heterojunction photovoltaic devices 
based on amorphous/crystalline interfaces.

\begin{acknowledgments}
The work was supported by U.S. National Science Foundation 
under Grants No. DMR 1507166 and No. DMR 1507118. The authors 
thank Profs.\;Gerard Barkema (Utrecht, The Netherlands) and 
Normand Mousseau (Quebec, Canada) for providing their 
modified WWW code.  The authors greatly appreciate the 
discussions with Profs.\;Stephen Elliott (Cambridge, UK) 
and David Drabold (Athens, Ohio).  One of us (RAF) 
acknowledges the Texas Advanced Computing Center (TACC) at 
The University of Texas at Austin for providing HPC resources 
that have contributed to the results reported in this work. 
\end{acknowledgments}


\begin{thebibliography}{64}
\bibitem [{par()}]{para}%
  \BibitemOpen
  \href@noop {} {}\bibinfo {note} {The readers may note that paracrystalline
  models,\cite{Khare} i.e., CRN models containing small clusters of ordered Si
  atoms, were employed in the past to explain the origin of the
  intermediate-range order in {\asi}, observed in Fluctuation Electron
  Microscopy.\cite{Gibson} For an alternative explanation, based on voids in
  CRN models, see Refs.\,\onlinecite{fem_ohio} and
  \onlinecite{MT_60}.}\BibitemShut {Stop}%
\bibitem [{vib()}]{vib}%
  \BibitemOpen
  \href@noop {} {}\bibinfo {note} {Since the energy associated with vibrational
  excitations in solids, involving a fraction of an electron-volt, is
  significantly smaller than their electronic counterpart (with a few to
  several electron-volts), a small perturbation in the local atomic environment
  of atoms can be readily manifested in the vibrational spectrum of a solid. At
  high temperature, such effects can be considerably weakened or washed out due
  to thermal broadening.}\BibitemShut {Stop}%
\end{thebibliography}
%

\end {document}